\def\year{2022}\relax
\documentclass[letterpaper]{article} 
\usepackage{aaai22}  
\usepackage{times}  
\usepackage{helvet}  
\usepackage{courier}  
\usepackage[hyphens]{url}  
\usepackage{graphicx} 
\usepackage{natbib}  
\usepackage{caption} 
\DeclareCaptionStyle{ruled}{labelfont=normalfont,labelsep=colon,strut=off} 
\usepackage{algorithm}
\usepackage{algorithmic}
\usepackage{tabularx}
\usepackage{multirow}
\usepackage{multicol}

\usepackage{newfloat}
\usepackage{listings}
\floatstyle{ruled}
\newfloat{listing}{tb}{lst}{}
\floatname{listing}{Listing}

\title{A Weakly-Supervised Iterative Graph-Based Approach to \\Retrieve COVID-19 Misinformation Topics}
\author{
    Harry Wang,
    Sharath Chandra Guntuku
}
\affiliations{
    University of Pennsylvania,\\ Department of Computer and Information Science,\\
    Philadelphia, PA 19104, USA\\
    harrywan@seas.upenn.edu, 
    sharathg@seas.upenn.edu
}

\usepackage{bibentry}

\begin{document}

\maketitle

\begin{abstract}
The COVID-19 pandemic has been accompanied by an `infodemic' -- of accurate and inaccurate health information across social media. Detecting misinformation amidst dynamically changing information landscape is challenging; identifying relevant keywords and posts is arduous due to the large amount of human effort required to inspect the content and sources of posts. We aim to reduce the resource cost of this process by introducing a weakly-supervised iterative graph-based approach to detect keywords, topics, and themes related to misinformation, with a focus on COVID-19. Our approach can successfully detect specific topics from general misinformation-related seed words in a few seed texts. Our approach utilizes the BERT-based Word Graph Search (BWGS) algorithm that builds on context-based neural network embeddings for retrieving misinformation-related posts. We utilize Latent Dirichlet Allocation (LDA) topic modeling for obtaining misinformation-related themes from the texts returned by BWGS. Furthermore, we propose the BERT-based Multi-directional Word Graph Search (BMDWGS) algorithm that utilizes greater starting context information for misinformation extraction. In addition to a qualitative analysis of our approach, our quantitative analyses show that BWGS and BMDWGS are effective in extracting misinformation-related content compared to common baselines in low data resource settings. Extracting such content is useful for uncovering prevalent misconceptions and concerns and for facilitating precision public health messaging campaigns to improve health behaviors.
\end{abstract}

\section{Introduction}

Since the COVID-19 pandemic hit the world, there have been multiple myths and misconceptions regarding the virus and associated health behaviors \cite{Singh_Bansal}. Many of the sources and contents of this misinformation (ranging from false cures to government conspiracies) have been disseminated online in various forms \cite{Cinelli_Quattrociocchi, Kouzy_Jaoude, Siwakoti_Yadav}. Misinformation and myths, especially during a public health crisis, are harmful and can lead to increased death along with poor health outcomes. \cite{Dharawat_Lourentzou, Memon_Carley, Ojo_Guntuku}. Gauging public perceptions aids in identifying specific sources of misinformation and concerns that need to be addressed in a tailored manner. 

\begin{figure}[!t]
\includegraphics[width=\columnwidth]{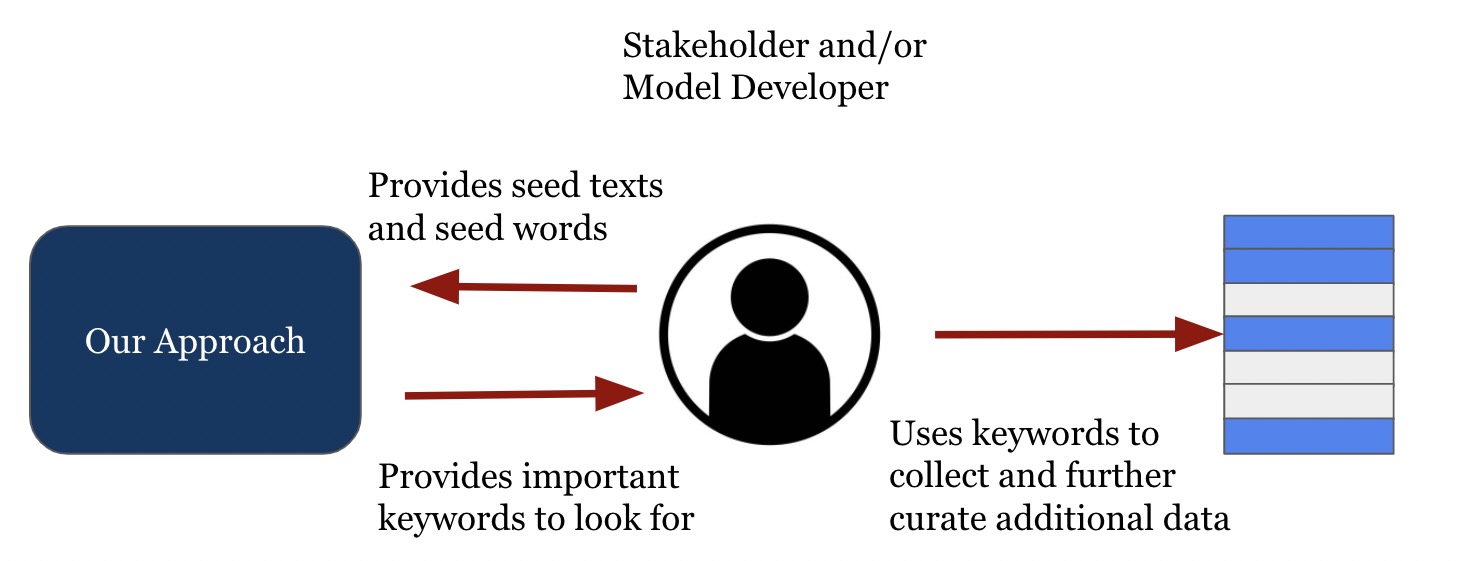}
\caption{Overview of our approach assisting stakeholders developing public health communication} \label{approach_overview}
\end{figure}

\begin{figure}[!b]
\includegraphics[width=.9\columnwidth]{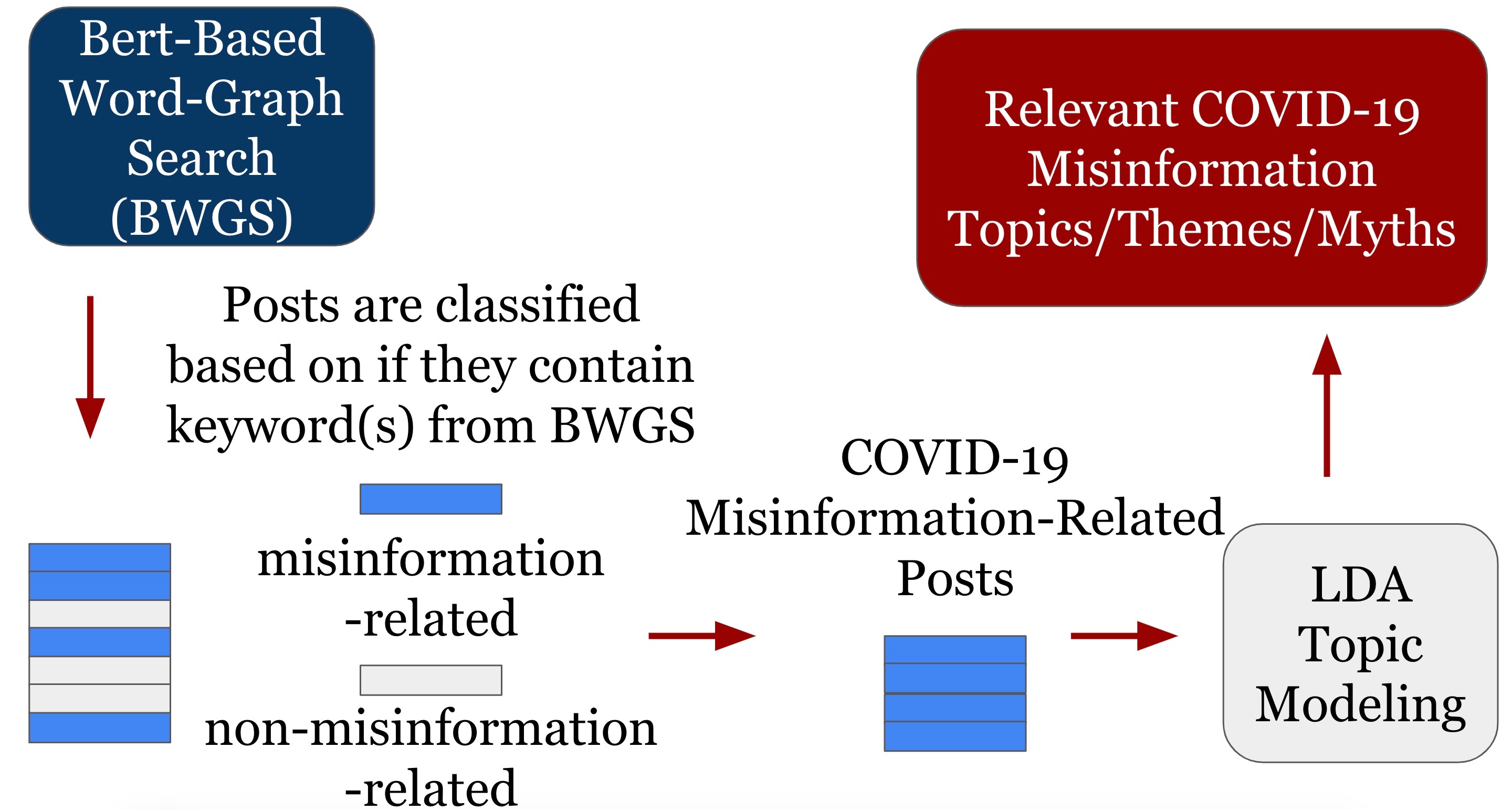}
\caption{Overview of our Weakly-Supervised Iterative Graph-Based Approach} \label{approach_overview}
\end{figure}

\begin{table*}[!t]
\small
\begin{tabular}{|p{2.5cm}|p{1.35cm}|p{8cm}|p{1.5cm}|p{2.15cm}|}
\hline
\textbf{Study}                                  & \textbf{Annotation} & \textbf{Dataset}                                                                                               & \textbf{Domain} & \textbf{Platforms}                     \\ \hline
\cite{Wang_Ma}                & Manual                   & Twitter: 7898 (fake), 6026 (real);Weibo:  4749 (fake), 4779 (real)                                              & Fake News       & Twitter, Weibo                           \\ \hline
\cite{Wang_Yang}              & Weakly Supervised        & Labeled:  2090 (fake), 2090 (real); Unlabeled: 22981                                                            & Fake News       & Wechat                                   \\ \hline
\cite{Vosoughi_Roy}           & URL matching             & 126,285 rumor cascades corresponding to 2,448 rumors                                                            & Rumors          & Twitter                                  \\ \hline
\cite{Dharawat_Lourentzou}    & Manual                   & 57,981 (real),  439 (possibly severe misinformation),  568 (highly severe), 1,851 (other), 447 (refutes/rebuts) & COVID-19        & Twitter                                  \\ \hline
\cite{Micallef_He}            & Manual                   & IFCN: 6,840 (false); annotated 2,400 tweets each for 5G conspiracy theories and fake cures                      & COVID-19        & Twitter                                  \\ \hline
\cite{Memon_Carley}           & Manual                   & 4573 tweets annotated into 17 categories                                                                        & COVID-19        & Twitter                                  \\ \hline
\cite{Cinelli_Quattrociocchi} & URL matching             & Gab: 3,778; Reddit: 10,084, YouTube: 351,786, Instagram: 1,328, Twitter: 356,448                                & COVID-19        & Twitter, Instagram, YouTube, Reddit, Gab \\ \hline
\cite{Shahi_Dirkson}          & Manual                   & 1274 (false),  226 (partially false)                                                                            & COVID-19        & Twitter                                  \\ \hline
\cite{Sharma_Seo}             & URL matching             & 150.8K Misinformation Source tweets                                                                             & COVID-19        & Twitter                                  \\ \hline
\cite{Brennen_Simon}          & Manual                   & 225 misinformation posts                                                                                        & COVID-19        & Twitter, Youtube, Facebook               \\ \hline
Our Approach                                    & Weakly Supervised        &   COVID-19 Keywords Tweets (26M) \cite{santosh2020detecting}; COVID19FN: 1596 (real), 1225 (fake) \cite{Agarwal}; CM-COVID-19~\cite{Memon_Carley}                                                                                                               & COVID-19        & Twitter, News Articles                   \\ \hline
\end{tabular}
\caption{\label{Related_Work_Analysis}Summary of related works including whether or not manual inspection of tweets or tweet sources was done, scale of data, whether the analysis was done on a COVID-19 dataset, and the social media platforms analyzed}
\end{table*}

Identifying and correcting misinformation is a challenging task given the different shapes and sizes it is propagated in~\cite{koltai2022addressing}. Given the large volume of social media and news article posts available, prior works have developed labeled COVID-19 datasets \cite{Agarwal, Memon_Carley}. These datasets label text based on whether misinformation is present and sometimes the type of misinformation that is present. Many approaches leverage posters' profiles and lists of credible sources to label posts as either misinformation or non-misinformation \cite{Vosoughi_Roy, Shahi_Dirkson}. Others rely on labeling either through professionals or crowd-sourcing\cite{Vraga_Bode, Van_Der_Meer_Jin, mujumdar2021hawkeye}. We reviewed several existing approaches to analyzing misinformation and summarize them in Table \ref{Related_Work_Analysis}. In this paper, we present a low-cost approach for retrieving COVID-19 misinformation related topics and themes with the goal of contributing to developing precision public health messaging campaigns to counter misinformation. 

One of the challenges in developing misinformation detection models is sifting through large amounts of unlabeled misinformation related sources to find relevant data to train models on. Our approach provides a complementary solution to this problem by generating a list of keywords, topics, and themes related to COVID-19 based on given seed words. The outputs of our approach can inform public health experts and model developers during the data collection process by providing keywords and topics to look for. Social media posts containing the keywords and topics generated from our approach can then be included in training data and/or further processed/curated. We also demonstrate the application of our approach on a down-stream misinformation detection task to provide a quantitative analysis of its performance. 

\begin{figure*}[!tp]
\centering
\includegraphics[width=.75\textwidth]{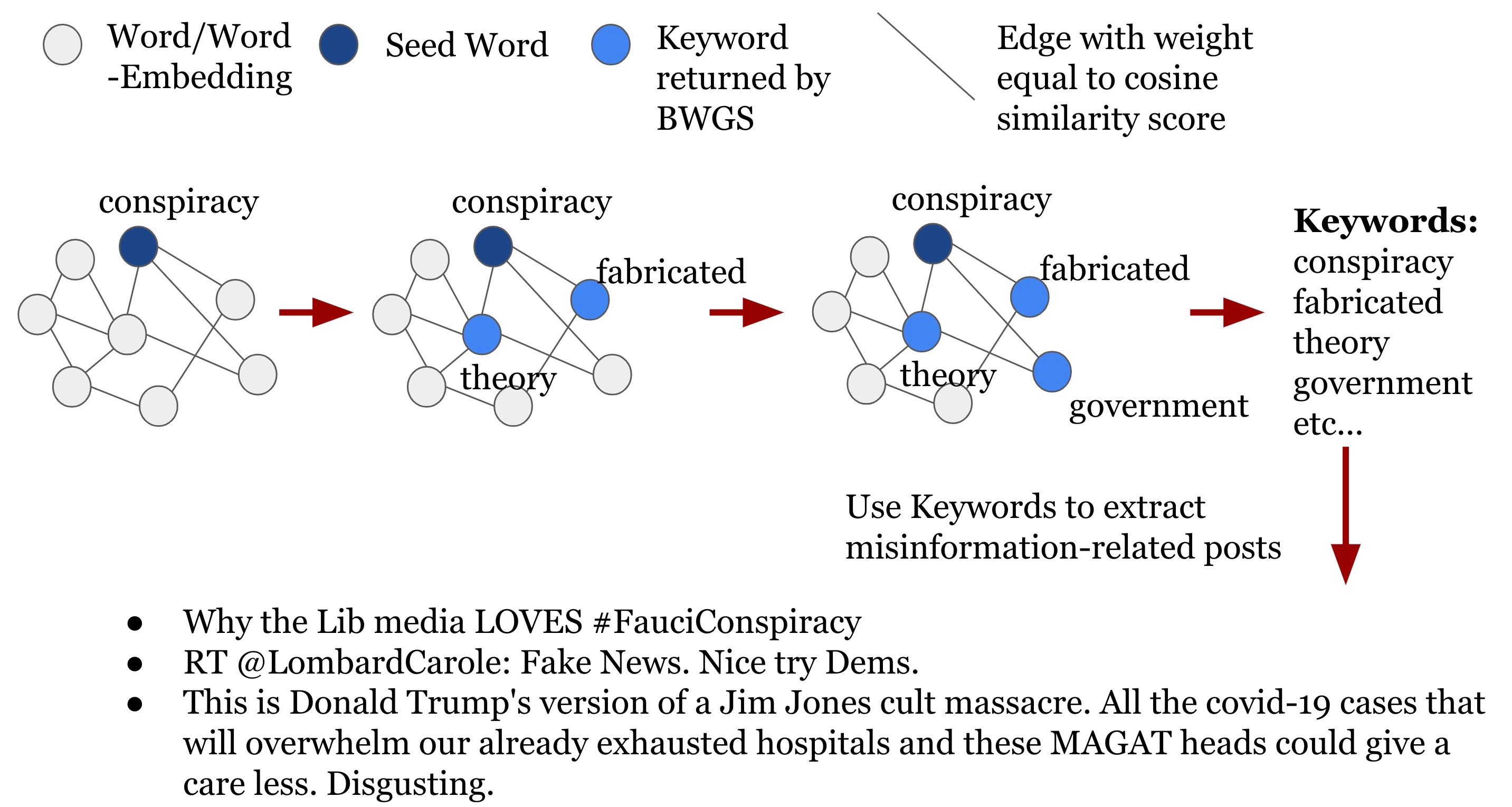}
\caption{Example run of Bert-Based Word Graph Search and sample results} \label{BWGS}
\end{figure*}

Another challenge of building misinformation detection models for a public health crisis during its early stages, such as the COVID-19 pandemic, is the lack of labels on pandemic-related misinformation. Fact-checking is a challenging task and can require large amounts of time and resources. While utilizing manual classification and/or supervision may achieve greater accuracy on a small amount of data, these approaches do not scale in terms of data volume and resources such as time and number of workers, especially in dynamically changing public health environments. Thus, it is important to develop a method to detect misinformation even in the presence of few labeled data points. 

Information and facts on pandemics can evolve over its course \cite{Brennen_Simon, Siwakoti_Yadav}. While manual labeling and annotation may provide a one-time solution to detecting misinformation, topics and sources of misinformation may change. New emergencies may also arise, rendering old labels obsolete. Thus, it is valuable to develop an automated, iterative method that requires significantly less human labor and manual inspection~\cite{Wang_Ma, Wang_Yang}. This can be achieved by decreasing the amount of supervision necessary for extracting misinformation topics and can make the process of detecting misinformation more repeatable at a lower cost. We achieve this by building upon BERT-Based Word Graph Search (BWGS) \cite{Santosh_Schwartz} to propose Bert-Based Multi-directional Word Graph Search (BMDWGS) in conjunction with Latent Dirichlet allocation (LDA) topic modeling. These results can help medical policy makers and public health practitioners gauge public perception towards the pandemic situation, thus creating more targeted campaigns to inform health behaviors and correct misinformation. Complementary to previous works, our approach does not require scraping a list of pre-defined reliable/non-reliable social-media/news article posters. It also does not need manual labeling and relies solely on few raw text data from social media or news posts for weak-supervision. We combine all of these characteristics and test them in the context of COVID-19 misinformation.

Our approach can serve as both a means to retrieve important COVID-19 misinformation keywords, topics, and themes and a method of classifying if a social media post is related to misinformation. We compare our approach to standard classification models in Natural-Language-Processing to provide a quantitative analysis. However, we do not directly compare our approach to the ones in \ref{Related_Work_Analysis} because our approach is designed for keyword and topic retrieval to inform data-collection processes for downstream tasks, rather than directly performing a downstream classification task.

In this work, we define a ``misinformation-related'' text as a piece of text that mentions a source of misinformation regarding COVID-19 but do not always directly endorse the statement of misinformation. An example of such a tweet would be the following: \textit{``Donald Trump's Closing Message to Americans: Ignore COVID-19 https://t.co/Iz1UK4cq4N Maybe the idiot needs to go to the families of those with covid and to families of the 200, 000+ that have died. And tell them it is fake and over. He is so good at lying.''} Extracting these tweets is useful for compiling a list of topics related to misinformation so that public health experts know which misconceptions are prevalent amongst the public \cite{Brennen_Simon, Kouzy_Jaoude, Singh_Bansal}.

\section{Methodology}

\subsection{Approach Overview}
We run BERT-Based Word Graph Search (BWGS) ~\cite{Santosh_Schwartz} on a word-embedding graph to obtain keywords of interest. BWGS relies on using seed texts to construct the word-embedding graph and a seed word as a starting point for the search algorithm. Seed texts are obtained by finding texts that contain the desired seed word. In our work, we select 50 texts from a target language corpus per seed word for each run of BWGS. This simulates a setting where few labeled datapoints are available. We compare the performance of our approach vs. baselines in this low data resource setting. BWGS constructs a word graph based on the tweets in the training corpus. A visualization of an overview of our approach is shown in Figure \ref{approach_overview}. 

We also propose BERT-Based Multi-directional Word Graph Search (BMDWGS). This algorithm improves upon BWGS by utilizing multiple seed words to begin the search algorithm. The search expands in multiple directions from the selected seed words. This method incorporates more starting context and information to guide the search. Seed texts are obtained by finding texts that contain all of the desired seed words. We select 50 texts from a training corpus per seed word set for each run of BMDWGS.

\begin{figure*}[!t]
\centering
\includegraphics[width=.75\textwidth]{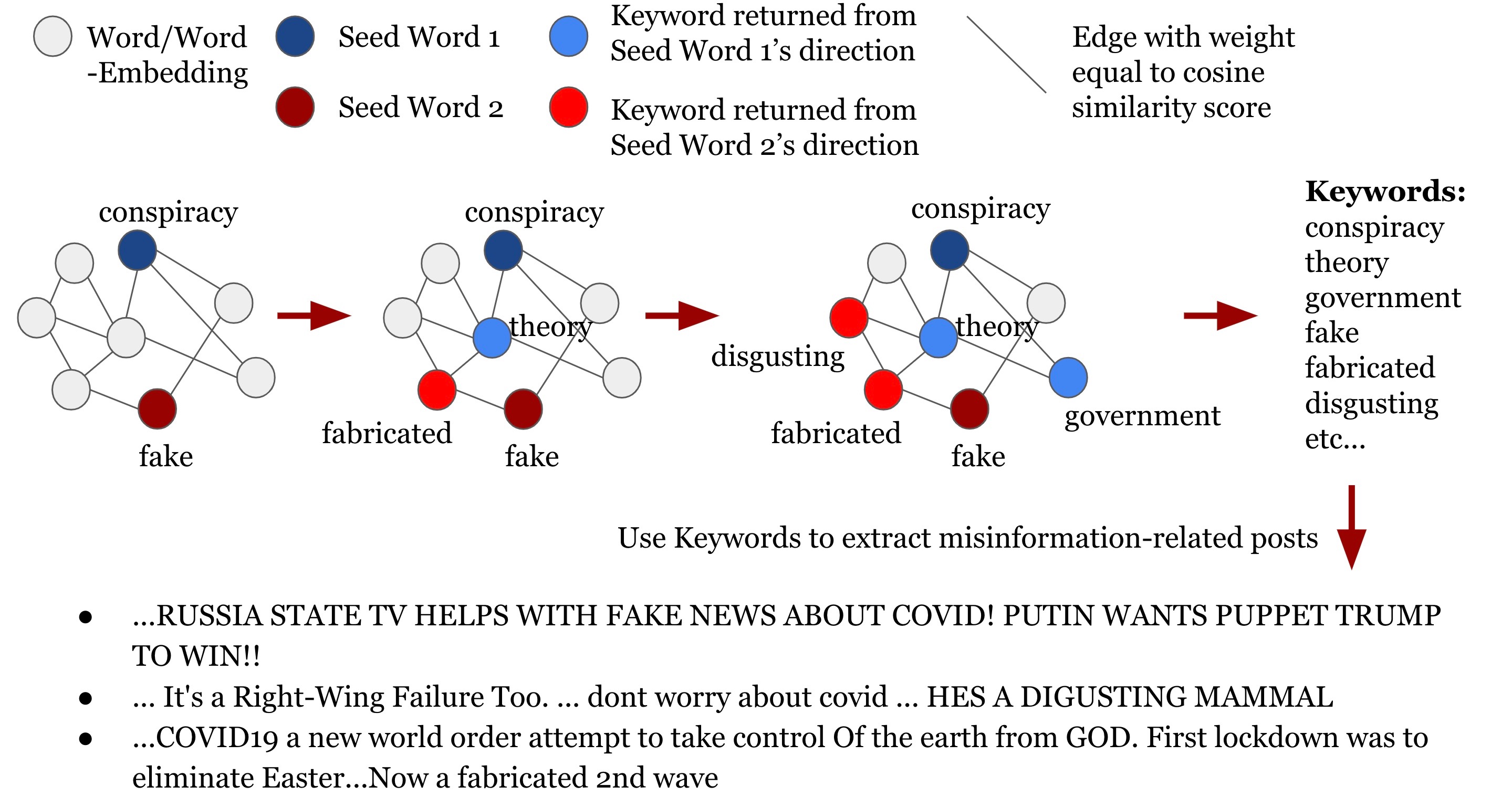}
\caption{Example run of Bert-Based Multi-directional Word Graph Search and sample results} \label{BMDWGS}
\end{figure*}

The next step of our approach involves returning all texts from the target language corpus containing the keywords returned from the BWGS/BMDWGS step. These texts are classified as ``misinformation-related''. We run Latent Dirichlet allocation (LDA) on the returned texts to obtain topics that are relevant in the misinformation-related texts. We then examine the top words corresponding to each topic to interpret relevant subjects within each topic. A visualization of results from this step is shown in Figure \ref{lda_results}. Compared to other methods, our seed word-based BWGS/BMDWGS retrieval and LDA Topic-Modeling make interpreting key features of misinformation-related texts easy and simple. While we use this approach to analyze COVID-19 related datasets, it is domain-agnostic \cite{Santosh_Schwartz}. 

We evaluate our approach on three different datasets, two of which have metrics to validate against. For both BWGS and BMDWGS, we utilize seed words that are commonly associated with misinformation, such as ``conspiracy'' and ``myth''. Texts containing these words are likely to mention a source of misinformation. We compare both BWGS and BMWGS against two baselines: a BERT Text Classification model, and a Logistic Regression model that been trained on a Bag-of-Words (BOW) feature vector. For the BERT Text Classification and Logistic Regression baselines, we up-sample from the minority-class of misinformation-related tweets to address label imbalance. In later sections, we show that our approach outperforms the baselines in the low linguistic resource settings. The source code used in this paper is available at \url{https://github.com/harryw1248/COVID_19_Misinformation_Weakly_Supervised_BWGS}


\begin{algorithm}[!t]
\caption{BERT-Based Word Graph Search (BWGS) \cite{Santosh_Schwartz}}
\begin{algorithmic}
\REQUIRE Initialized graph $\textbf{G}$, initialized queue $\textbf{Q}$, initialized Context Embedding $\textbf{CEmb}$, minimum similarity threshold $minSimThresh$, maximum depth $maxDepth$, top k-words $k$\\
\STATE Initialize empty list $keywords$
\STATE Initialize current depth $depth=0$

\STATE 1. Pop next word from $\textbf{Q}$, denoted by $t$. Initialize a new node in $\textbf{G}$ based on $t$ and set the node embedding to $\textbf{Emb}\{t\}$. Append $t$ to $keywords$
\STATE 2. Initialize the query embedding q as $q \leftarrow k \times CEmb+(1-k) \times Emb\{t\}$.
\STATE 3. Iterate through all tokens in the data. We compare the tokens' embeddings against the query embedding $q$. We drop all tokens with similarity less than the minimum similarity threshold $minSimThresh$.
\STATE 4. Select the top $k$ words that are most similar to $q$. Append these words to $\textbf{Q}$. Instantiate new nodes (if one does not already exist) for these words in $\textbf{G}$ and add outgoing edges from $t$ to these new nodes.
\STATE 5. If all words for given depth are explored, the top $k$ words corresponding to that depth are selected based on similarity to $\textbf{CEmb}$. Update the context embedding by averaging $\textbf{CEmb}$ with the representative embeddings of the selected words: $\textbf{CEmb} \leftarrow \frac{CEmb+\sum_{i=0}^m Emb\{x_i\}}{m+1}$. Increment $depth$ by 1.
\STATE 6. Terminate iterations when either $\textbf{Q}$ is empty or when $depth=maxDepth$. Otherwise, repeat from Step 1. If a stopping condition is met, return $keywords$.
\end{algorithmic}
\end{algorithm}

\begin{algorithm}[t!]
\caption{BERT-based Multi-directional Word Graph Search (BMDWGS)}
\small

\begin{algorithmic} \label{BMDWGS_algo}
\REQUIRE minimum similarity threshold $minSimThresh$, maximum depth $maxDepth$, top k-words $k$, list of seed words $seedWords$\\
\STATE Initialize empty list $keywordsUnion$

\FOR {$seedWord$ in $seedWords$}
    \STATE 1. Initialize $\textbf{G}$, queue $\textbf{Q}$, Context Embedding $\textbf{CEmb}$ using $seedWord$
    \STATE 2. Run BWGS with $\textbf{G}$, $\textbf{Q}$, $\textbf{CEmb}$, $minSimThresh$, $maxDepth$, $k$ to obtain $keywords$
    \STATE 3. $keywordsUnion\leftarrow keywordsUnion\cup keywords$
\ENDFOR
\STATE return $keywordsUnion$
\end{algorithmic}
\end{algorithm}

\subsection{BERT-based Word Graph Search (BWGS)}

We leverage BERT, a state-of-the-art Deep Bidirectional Transformers-based language model for constructing word-embeddings \cite{Devlin_Chang}. We fine-tune a bert-base-uncased model using a masked language modeling protocol with the following parameters: Num. Epochs: 5, Learning Rate: 1e-5, Batch Size: 8, Max Sequence Length: 128 \cite{Santosh_Schwartz}. We obtained 768-dimensional word-embeddings from this model by summing the hidden states of the last 4 BERT layers. HuggingFace library is used for loading pre-trained model weights as well as Language Model training \cite{Wolf_Debut}. 

Using the word-embeddings, we construct a word-graph where node locations are determined by the word-embeddings. Nodes in this graph represent words, and edge weights represent the degree of similarity between two words. A search is performed through this word-graph, where tokens have their embedding compared against a query embedding $q$. If the cosine similarity between a token's embedding and the query embedding $q$ is below the minimum similarity threshold, $minSimThresh$, the token is dropped and will not be explored further. 

Previous work has found success in using a “Context Embedding,” $\textbf{CEmb}$ to search through these word graphs. These Context Embeddings are constructed by using a seed word for initialization and integrating information from other words over iterations, following the formula $\textbf{CEmb} \leftarrow \frac{CEmb+\sum_{i=0}^m Emb\{x_i\}}{m+1}$ where $m$ corresponds to the top $m$ words at a certain depth in the search process. This creates a more robust and comprehensive representation of the desired context. The hyperparameter, $k$ allows us to control the number of tokens that are considered at each depth in the search process. Thus, the full formula for the query embedding $q$ is $q \leftarrow k \times CEmb+(1-k) \times Emb\{t\}$ where $t$ is the current token being explored. A visualization for BWGS is shown in Figure \ref{BWGS} \cite{Santosh_Schwartz}.


\subsection{BERT-Based Multi-directional Word Graph Search (BMDWGS)}

BERT-Based Multi-directional Word Graph Search improves upon BWGS by incorporating more starting context information into the search process. This is done by using multiple seed words. As a result, BMDWGS is able to retrieve more relevant keywords than BWGS given the same $maxDepth$ parameter value. We show in later results that using this technique results in more robust performance, particularly with respect to recall scores. A visualization for BMDWGS is shown in Figure \ref{BMDWGS}. The specific steps used in the BMDWGS algorithm are found in Algorithm \ref{BMDWGS_algo}.

\section{Analysis}

We validate our approach on three different datasets: a COVID-19 Keywords Tweet Dataset \cite{Santosh_Schwartz}, the COVID19FN Dataset \cite{Agarwal}, and the Carnegie Mellon COVID-19 Tweet Dataset \cite{Memon_Carley}. We utilize the COVID-19 Keywords Tweet Dataset to demonstrate the qualitative nature of the results returned from our weakly supervised approach. This includes misinformation-related keywords and top words in misinformation-related topics. The data from the COVID-19 Keywords Tweet Dataset also contains more recent tweets. 
We then use the model trained on our COVID-19 Keywords Tweet Dataset to predict labels from the COVID19FN and the Carnegie Mellon COVID-19 Tweet Dataset. The COVID19FN dataset contains news articles related to COVID-19 misinformation and allows us to validate our approach and obtain quantitative results in terms of precision, recall, and f1-score when labels are presented in a binary misinformation/non-misinformation format. It also allows us to examine how well our method can transfer from a Tweet-domain to a news-article domain. We also utilized the Carnegie Mellon COVID-19 Tweet Dataset to validate our approach on a dataset with non-binary labels.

\subsection{COVID-19 Keywords Tweet Dataset}
\subsubsection{Data and Model Hyperparameters}
We utilize tweets collected in November 2020 related to COVID-19. This dataset was also used in other works related to studying emerging topics, symptoms, and varying health behaviors during COVID-19 \cite{vanexplaining, zamani2020understanding, santosh2020detecting}. We select a random subset of 10M tweets from over 26M total tweets. After filtering for English tweets, the dataset yielded over 5.5M tweets. These tweets serve as our training corpus for finding keywords in all of our experiments. For running BWGS, we used $seedWord: \textnormal{``conspiracy''}$, $similarityThreshold: 0.4$, $maxDepth: 2$, and top $k: 4$. We then run LDA using the MALLET implementation in DLATK~\cite{schwartz2017dlatk} with $numTopics$: 25 on these tweets to examine topics. We select ``conspiracy" as the seed word because it is commonly associated with the context of misinformation. Additionally, many COVID-19 misinformation theories have been conspiracy-themed, such as relating COVID to government labs or certain political agendas. In subsequent analyses, we utilize other seed words, such as ``claim" and ``hoax," as these words are commonly associated with misinformation and can be connected to a broader set of topics related to misinformation. Furthermore, during the early stages of a public health crisis, specific misinformation topics may be unknown. Thus, data collection could rely on general words related to the context of misinformation. 
\begin{figure}[!t]
\includegraphics[width=\columnwidth]{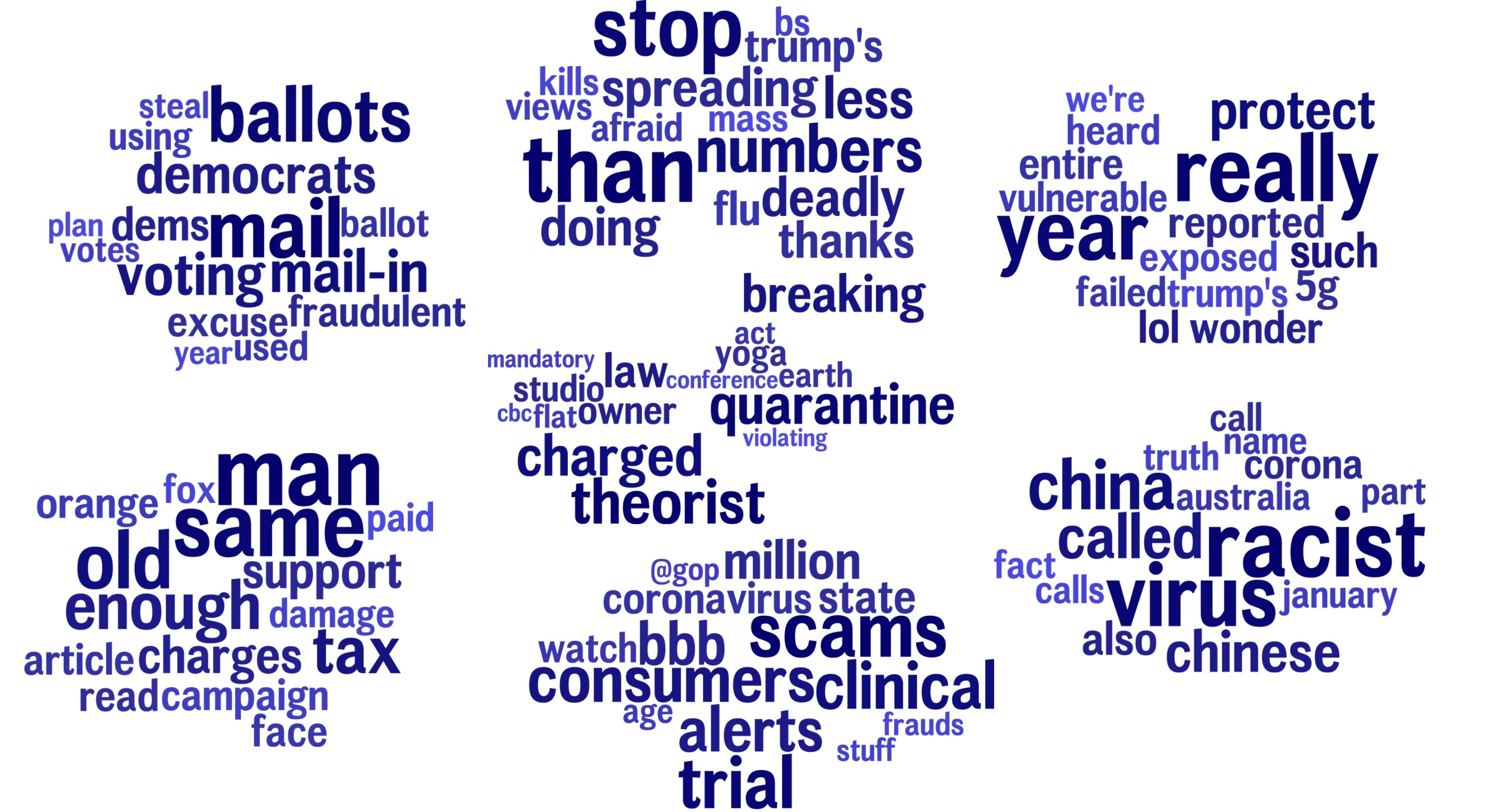}
\caption{Top Words in Sample Topics. The LDA step of our approach produces topic clusters that can be used for identifying top misinformation themes.} \label{lda_results}
\end{figure}

\begin{table*}[!t]
\centering
\small
\begin{tabular}{|l|l|l|l|l|l|l|}
\hline
\textbf{Baseline} & & & & Precision & Recall & F1-Score\\
\hline
BERT Text Classification & & & & 0.396 & 0.832 & 0.537\\
\hline
Logistic Regression BOW & & & & 0.171 & 0.0699 & 0.0992\\
\hline
\textbf{BWGS} & & & & & & \\
\hline
Seed Word(s) & Similarity Threshold & Max Depth & Top k &  & & \\
\hline
``conspiracy'' & 0.4 & 2 & 4 & 0.912 & 0.288 & 0.438\\
\hline
``claim'' & 0.4 & 2 & 4 & 0.813 & 0.820 & 0.817\\
\hline
``hoax'' & 0.4 & 2 & 4 & 0.936 & 0.323 & 0.480\\
\hline
``false'' & 0.4 & 2 & 4 & 0.835 & 0.788 & 0.811\\
\hline
``myth'' & 0.4 & 2 & 4 & 0.872 & 0.815 & 0.843\\
\hline
``myth'' & 0.4 & 3 & 4 & 0.752 & 0.919 & 0.827\\
\hline
``myth'' & 0.4 & 4 & 4 & 0.720 & 0.972 & 0.828\\
\hline
``myth'' & 0.3 & 2 & 4 & 0.815 & 0.921 & \textbf{0.865}\\
\hline
``myth'' & 0.5 & 2 & 4 & 0.877 & 0.814 & 0.844\\
\hline
``myth'' & 0.4 & 2 & 5 & 0.758 & 0.597 & 0.668\\
\hline
``myth'' & 0.4 & 2 & 6 & 0.658 & 0.987 & 0.790\\
\hline
\textbf{BMDWGS} & & & & & & \\
\hline
[``conspiracy'',``claim''] & 0.4 & 2 & 4 & 0.837 & 0.822 & 0.829\\
\hline
[``myth'',``claim''] & 0.4 & 2 & 4 & 0.640 & 0.987 & 0.777\\
\hline
[``myth'',``false''] & 0.4 & 2 & 4 & 0.565 & 0.999 & 0.722\\
\hline
[``conspiracy'',``myth''] & 0.4 & 2 & 4 & 0.834 & 0.795 & 0.814\\
\hline
[``conspiracy'',``false''] & 0.4 & 2 & 4 & 0.814 & 0.897 & 0.853\\
\hline
[``conspiracy'',``false''] & 0.4 & 3 & 4 & 0.768 & 0.918 & 0.836\\
\hline
[``conspiracy'',``false''] & 0.4 & 4 & 4 & 0.709 & 0.960 & 0.816\\
\hline
[``conspiracy'',``false''] & 0.3 & 2 & 4 & 0.815 & 0.896 & \textbf{0.854}\\
\hline
[``conspiracy'',``false''] & 0.5 & 2 & 4 & 0.824 & 0.879 & 0.850\\
\hline
[``conspiracy'',``false''] & 0.4 & 2 & 5 & 0.565 & 0.998 & 0.722\\
\hline
[``conspiracy'',``false''] & 0.4 & 2 & 6 & 0.565 & 1.00 & 0.722\\
\hline
\end{tabular}
\caption{\label{Covid_19fn}
BWGS and BMDWGS Results on COVID19FN Dataset based on different hyperparameter sets; highest f1-scores are in bold
}
\end{table*}

\subsubsection{Findings}
The COVID-19 Misinformation Keyword tokens returned by BWGS include: ``fake'', ``theories'', ``racist'', ``theory'', ``fabricated'', ``fraud'', ``theorists'', ``\#\#cies'', ``horrible'', ``hatred'', ``nationalist'', ``disgusting'', ``venture'', ``conspiracy.'' 

According to our approach, 0.6717\% of tweets (37026 out of the over 5.5M tweets) were labeled as COVID misinformation-related tweets. Due to the lack of labels for this dataset, we relied on the manual inspection of 100 tweets classified by our approach as misinformation-related and 100 tweets classified by our approach as non-misinformation-related. Upon inspection, we found that 84 of the 100 tweets classified as misinformation-related were indeed true positives and that 91 of the 100 tweets classified as non-misinformation-related were indeed true negatives.

LDA topics associated with misinformation and myths were related to: Trump, politics, government conspiracies, 5G, China, vaccines, lockdowns, quarantine, COVID-19 affecting voter-fraud, racism, etc. We present the top words related to sample topics in Figure \ref{lda_results}. While we started off with the general misinformation-related word ``conspiracy,'' our approach was able to retrieve a wide-range of specific misinformation-related topics due to shared context.

\subsection{COVID19FN Dataset}

\subsubsection{Data and Model Hyperparameters}

We validate our approach using the COVID19FN dataset \cite{Agarwal}. This dataset consists of labelled news articles of misinformation spread during the COVID-19 infodemic. It contains approximately 2800 news articles, real and fake, scraped from Poynter and other fact-checking sites. This dataset has binary labels that indicate whether an article is related to misinformation.

We experiment with different values for the hyperparameters in both BWGS and BMDWGS. For each set of hyperparameters, we measure precision, recall, and f1-score, three commonly used metrics in Natural Language Processing and Information Retrieval \cite{Goutte_Gaussier}.

\subsubsection{Findings}


We present our findings on this dataset in Table \ref{Covid_19fn}. We observe that most of the BWGS trials outperformed both the BERT Text Classification and Logistic Regression baselines. Of the 11 BWGS trials, 9 outperformed both of the baselines. BWGS with ``myth'' as the seed word, $similarityThreshold=0.3$, $maxDepth=2$, and top $k=4$ yielded an f1-score of 0.865, the highest of all the BWGS trials on this dataset. It also yielded a recall score of 0.921 and a precision score of 0.815. It achieved 32.8\% higher f1, 8.9\% higher recall, and 41.9\% higher precision compared to the BERT Text Classification baseline. The Logistic Regression Baseline did not achieve above 0.50 for any of the three metrics. 

We observe that all of the BMDWGS trials outperformed both the BERT Text Classification and Logistic Regression baselines. BMDWGS with [``conspiracy'',``false''] as the seed word set, $similarityThreshold=0.3$, $maxDepth=2$, and top $k=4$ yielded an f1-score of 0.854, the highest of all the BMDWGS trials on this dataset. It also yielded a recall score of 0.896 and a precision score of 0.815. It achieved 31.7\% higher f1, 6.4\% higher recall, and 41.9\% higher precision compared to the BERT Text Classification baseline.

\subsubsection{BWGS}

To analyze the effect of varying hyperparameters, we experimented with five different seed words with fixed $similarityThreshold$, $maxDepth$, and top $k$ values. We then selected the seed word that achieved the highest f1-score to observe the effects of varying $similarityThreshold$, $maxDepth$, and top $k$ values. Further experimentation that involves trying every hyperparameter combination could have been done. However, we were primarily interested in examining the effects of varying $similarityThreshold$, $maxDepth$, and top $k$ values with respect to a certain seed word. 

We select ``myth'' as the seed word for testing different $similarityThreshold$, $maxDepth$, and top $k$ values as it yielded the highest f1-score (0.843) in our first 5 trials. We observe that increasing $maxDepth$ produced increasing trends in recall and decreasing trends in precision. This is expected, due to the fact that more keywords are being returned, increasing the number of articles retrieved, thus leading to more true positives and false positives. A similar trend can be observed with increasing the top $k$ value. We found no definitive pattern with increasing $similarityThreshold$. Similar to many other information retrieval studies, we observe a general trade-off between precision and recall. However, multiple sets of hyperparameters were able to achieve greater than 0.800 in precision, recall, and f1-score.

\subsubsection{BMDWGS} \label{BMDWGS_FN}

Our analyses with BMDWGS are similar to those with BWGS except that five different seed word sets are used instead.


We select [``conspiracy'',``false''] as the seed word set for testing different $similarityThreshold$, $maxDepth$, and top $k$ values as it yielded the highest f1-score (0.853) in our first 5 trials. Interestingly, the two seed words from the first 5 BWGS trials (``myth'' and ``claim'') that yielded the highest f1-score did not form the seed word set that produced the highest f1-score using BMDWGS. For BMDWGS, we observe similar patterns in precision and recall varying with $similarityThreshold$, $maxDepth$, and top $k$ values. 

BMDWGS achieves 17.0\% higher average recall (0.920 versus 0.750 ), but has 9.20\% lower average precision (0.722 versus 0.814). This is likely due to the fact that the $keywordUnion$ returned by BMDWGS has a greater cardinality than than the $keywords$ set returned by BWGS given a fixed hyperparameter set, leading to more texts being returned. We also observe that overall, BMDWGS had more robust performance on this dataset -- the standard deviation of f1-scores and recall scores across BMDWGS hyperparameter sets was 9.20\% lower (0.0520 versus 0.144) and 16.4\% lower (0.0693 versus 0.233) respectively, while the standard deviation of precision scores was only 2.99\% higher (0.111 versus 0.0811). This is likely due to the fact that utilizing multiple seed words allows BMDWGS to take more contextual information into account, leading to a more comprehensive language representation. BMDWGS is also less likely to be sensitive to the effects of picking one particular seed word. 

\subsection{Carnegie Mellon COVID-19 Tweet Dataset}
\subsubsection{Data and Model Hyperparameters}
We also validated our approach using the Carnegie Mellon COVID Tweet Dataset because it contained multiple classes of labels (beyond binary) and multiple labels per tweet \cite{Memon_Carley}. The dataset contains 4573 annotated tweets and 17 classes of labels. Our experiments on the COVID-19 dataset suggest that our approach is successful in detecting conspiracy-based misinformation. Thus, we aimed to retrieve tweets labeled as ``conspiracy'' and ``calling out or corrections,'' as they provide insights into perceptions towards conspiracies related to COVID. We experimented with several sets of hyperparameters for both BWGS and BMDWGS. We found that BMDWGS outperformed BWGS in terms of average f1-score and found the following hyperparameter set to yield the highest f1-score of our experiments: $seedWordSet: \textnormal{[``conspiracy'', ``myth'']}$, $similarityThreshold: 0.45$, $maxDepth: 6$, and top $k: 6$. We found the following hyperparameter set to yield the highest f1-score for BWGS: $seedWord: \textnormal{``conspiracy''}$, $similarityThreshold: 0.55$, $maxDepth: 4$, and top $k: 4$.

\subsubsection{Findings}

We present our findings on this dataset in Tables \ref{CMU_findings_Baselines} (baseline performance), \ref{CMU_findings_BWGS} (BWGS performance), and \ref{CMU_findings_BMDWGS} (BMDWGS performance). BMDWGS achieved higher f1-score and recall than BWGS while BWGS achieved higher precision. This is consistent with the explanations and observations from the experiments run on the COVID19FN dataset.

\begin{table}[!t]
\centering
\small
\begin{tabular}{|l|l|l|l|}
\hline
Baseline & Precision & Recall & F1\\
\hline
BERT Text classification & 0.478 & 0.933 & 0.632\\
\hline
Logistic Regression BOW & 0.333 & 0.00173 & 0.00344 \\
\hline
\end{tabular}
\caption{\label{CMU_findings_Baselines}
BERT Text Classification and Logistic Regression Baseline Results on Carnegie Mellon COVID Tweet Dataset
}
\end{table}

\begin{table}[!t]
\centering
\small
\begin{tabular}{|l|p{5.5cm}|}
\hline
Keyword tokens & ``fake'', ``conspiracy'', ``theory'', ``fabricated'', ``fraud'', ``\#\#s'', ``\#\#cies'', ``\#\#ted'', ``\#\#ed'',  ``corruption'', ``folks'', ``\#\#ers'', ``theorists'', ``\#\#led'', ``\#\#s''\\
\hline
Precision & 0.745 \\
\hline
Recall & 0.202 \\
\hline
F1-Score & 0.317 \\
\hline
\end{tabular}
\caption{\label{CMU_findings_BWGS}
BWGS Results on Carnegie Mellon COVID Tweet Dataset, including COVID-19 Misinformation Keyword tokens returned by BWGS
}
\end{table}

\begin{table}[!t]
\centering
\small
\begin{tabular}{|l|p{5.5cm}|}
\hline
Keyword tokens & ``false'', ``myths'', ``theory'',  ``theorists'', ``\#\#ation'', ``mis'',  ``\#\#cies'', ``skepticism'', ``idea'', ``belief'', ``\#\#form'', ``published'',  ``con'', ``discontent'', ``device'',  ``paranoia'', ``bust'', ``control'',  ``hate'', ``another'',  ``conspiracy'', ``\#\#ation'',  ``\#\#cies'', ``nonsense'',  ``paranoia'', ``nonsense''\\
\hline
Precision & 0.471 \\
\hline
Recall & 0.502 \\
\hline
F1-Score & 0.486 \\
\hline
\end{tabular}
\caption{\label{CMU_findings_BMDWGS}
BMDWGS Results on Carnegie Mellon COVID Tweet Dataset, including COVID-19 Misinformation Keyword tokens returned by BMDWGS
}
\end{table}



We observe that our approach reveals certain limitations of keyword based retrieval methods. The fact that this dataset contained non-binary labels and multiple labels per tweet also made this retrieval task particularly challenging. 

False negatives included tweets that were labeled as ``conspiracy'' and/or ``calling out or corrections.'' We observe that many of these false negative tweets contained many generic words alongside one keyword that corresponded with misinformation such as ``bleach'' or ``cocaine.'' These keywords were not used by our approach as BWGS and BMDWGS were unable to find these keywords given the seed words. False positives included tweets from all other 15 label classes. Many of these tweets contained generic words that were listed as keywords by our approach such as ``discontent'' and ``hate.'' Since words similar to these co-occur often with the seed words, BWGS and BMDWGS returned these as keywords. Both BWGS and BMDWGS outperformed the Logistic Regression baseline in terms of precision, recall, and f1-score. BERT Text classification outperformed BWGS and BMDWGS in terms of recall and f1-score. BWGS outperformed BERT Text Classification in terms of precision. 

\section{Discussion}

Our approach combines techniques from state-of-the-art neural netowrk based embedding models, algorithmic search, and topic modeling to successfully identify key themes of misinformation regarding COVID-19. It identified misinformation related keywords, topics, and themes related to: Trump, politicians, government conspiracies, vaccines, lockdowns, quarantine, COVID-19 affecting voter-fraud, racism, etc from general misinformation related seedwords. Knowing these keywords can inform developers of downstream models and public health officials about what topics to collect data on to improve the performance of downstream systems. 

During public health emergencies, information and misinformation are likely to propagate through populations at a rapid pace. Myths and conspiracies have the potential to cause irrational or scientifically unsound decisions by individuals. These actions can pose health hazards and could lead to increased mortality. Thus, it is important to be able to quickly detect these sources of information so that public health officials have the ability to correct (mis)beliefs. Given our access to social media and the large quantity of data present, manual annotation can prove to be cost and resource intensive to implement during certain situations. The state of emergencies also evolve throughout time and new trends and sources of misinformation may arise. Thus, an easily automated and scalable means of finding key themes of misinformation is useful. 

In addition to assisting a ``human-in-the-loop'' that is responsible for collecting data and training models, our Weakly-Supervised Iterative Graph-Based Approach also has useful characteristics when directly used for downstream misinformation detection tasks.  It does not require manual labeling of datasets to be able to find social media texts related to misinformation. We also improve upon the existing BWGS method by introducing a novel BMDWGS algorithm to utilize greater starting context. The fact that our methods only require raw text input without manual annotations allows them to be easily automated, complementing and potentially making existing fact-checking efforts more efficient. 

While BMDWGS improves upon BWGS by using more seed words for better contextual information, an even greater amount of context can be achieved. An n-gram graph search could be utilized that uses an n-gram as a seed rather than a single word. An n-gram graph could be constructed by embedding n-grams present in the corpus. This could lead to greater performance because generating n-grams would allow BWGS and BMDWGS to encode greater context information. Both methods are able to produce keywords that can be used to extract misinformation related texts. Expanding them to n-grams would allow them to go beyond this scope. Our current approach achieved relatively high performance on datasets with binary labels, but further investigation is needed to improve performance on more complex datasets with several classes. Further exploration could also be done on analyzing misinformation regarding events other COVID-19. While this work specifically analyzes COVID-related misinformation given current events, its methodology is meant to be domain-agnostic. 

Our approach also has applications in Information Retrieval and Recommender Systems, particularly with Query Expansion. Query expansion is the process of supplementing additional terms or phrases to the original query to improve the retrieval performance \cite{Imran_Sharan}. Our approach can be used to automatically search for related terms to a query based on learned context and embeddings in low data resource settings, alleviating the data cold-start problem in certain scenarios. The ability to search for related terms is also applicable in psychology research, where word lists such as LIWC are useful for analyzing participants' mental states when analyzing words used in text messages \cite{Tausczik_Pennebaker}. 

\section{Conclusion}
We introduce a weakly-supervised iterative graph-based approach for detecting topics related to COVID-19 misinformation. Our approach successfully extracted many key topics related to misinformation and produced lists of important keywords while using little training data. The novel BMDWGS algorithm we propose also shows greater performance in terms of f1-score, recall, and overall robustness to hyperparameter tuning. Further work could explore the effects of using our approach for different domains, as well as expanding upon its methodology. The work presented in this paper sets the groundwork for developing low data-resource,  weakly-supervised, graph-based approaches for misinformation detection.

\section{Acknowledgements}
This project was supported by Penn Global and the India Research and Engagement Fund.

\bibliography{aaai22.bib}
\bibliographystyle{aaai22.sty}
\end{document}